\documentclass[aps,prl,twocolumn,groupedaddress,showpacs]{revtex4}

\usepackage{graphicx}
\usepackage{epsfig}

\begin{document}

\DeclareGraphicsExtensions{.eps,.EPS,.jpg,.bmp}


\title{Laser controlled tunneling in a vertical optical lattice}


\author{Q. Beaufils, G. Tackmann, X. Wang, B. Pelle, S. Pelisson, P. Wolf and F. Pereira dos Santos}

\email[]{franck.pereira@obspm.fr}
\affiliation{LNE-SYRTE, UMR 8630 CNRS, Observatoire de Paris,
UPMC, 61 avenue de l'Observatoire, 75014 Paris, FRANCE}


\date{\today}

\begin{abstract}
Raman laser pulses are used to induce coherent tunnelling between neighbouring sites of a vertical 1D optical lattice.
Such tunneling occurs when the detuning of a probe laser from the atomic transition frequency matches multiples of the Bloch frequency,
allowing for a spectroscopic control of the coupling between Wannier Stark (WS) states. In particular, we prepare coherent superpositions of WS states of adjacent sites,
and investigate the coherence time of these superpositions by realizing a spatial interferometer. This scheme provides a powerful tool for coherent manipulation of external degrees of freedom of cold atoms, which is a key issue for quantum information processing.

\end{abstract}


\maketitle



Trapping and manipulating cold neutral atoms in an optical lattice offers high controllability and robust quantum coherence properties, which makes it an attractive system for many applications such as quantum simulation of solid state systems \cite{cirac}, metrology \cite{latticeclock,hsurm}, and quantum information processing (QIP) \cite{qip}. One key issue in this context is the possibility to coherently control the atoms internal and external degrees of freedom. Combined with the possibility to address single sites \cite{single}, this allows for the realization of quantum logic operations \cite{ql}.

Atom transport control in an optical lattice has been previously
reported using microwave fields \cite{microwave}, frequency, phase
and amplitude modulation techniques \cite{salomon,raizen,tino}, or
an adiabatic change of the trapping potential \cite{Bloch,Porto}.
In this work, we demonstrate coherent laser induced
tunneling of cold atoms between neighboring sites of an optical
lattice. In contrast with most previous approaches, our technique
doesn't require any modification of the trapping potential. It
allows an unprecedent control of the atom's external degrees of
freedom (displacing the atoms by 1 to 9 lattice periods in this
work) in a system showing good coherence properties (up to $1$ s).

Our system consists in laser-cooled $^{87}$Rb atoms in the first
band of a vertical one-dimensional optical lattice. Due to earth
gravity, the ground energy levels of the lattice are shifted out
of resonance. For a sufficiently large lattice depth $U_{l}$,
tunneling is highly reduced, leading to a ladder of localized
Wannier-Stark (WS) eigenstates separated by the Bloch frequency
$\nu_{B}=m_{a}g\lambda_{l}/2h$. Here, $m_{a}$ is the atomic mass,
$g$ is the gravity acceleration, $\lambda_{l}/2$ is the distance
between two adjacent lattice sites, and $h$ is the Planck constant.
The WS states $\left|W_{m}\right\rangle$ are indexed by the
discrete quantum number $m$ characterizing the well containing the
center of the wave function $\left\langle
x\right.\left|W_{m}\right\rangle$.

\begin{figure}
    \centering
        \includegraphics[scale=0.6]{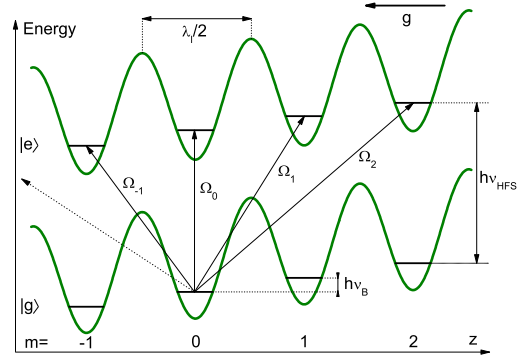}
   
    \caption{Atoms in the first band of the lattice form a Wannier-Stark ladder of
eigenstates. The Raman probe laser couples the ground to the
excited hyperfine level in the different WS states separated by
the Bloch frenquency. }
\label{fig:energy}
\end{figure}

We use counterpropagating Raman beams to drive coherent
transitions between the ground and excited hyperfine levels
$\left|g\right\rangle
=\left|5^{2}S_{1/2},F=1,m_{F}=0\right\rangle$ and
$\left|e\right\rangle
=\left|5^{2}S_{1/2},F=2,m_{F}=0\right\rangle$. Such a transition 
implies a momentum transfer of
$k_{eff}=k_{1}+k_{2}\approx 4\pi/(780$ nm) that couples the WS
states either in the same well or in neighboring wells, with a
coupling strength proportional to $\left\langle
W_{m}\right|e^{ik_{eff}x}\left|W_{m\pm \Delta m}\right\rangle$.
Fourier-limited widths of the resonances over excitation times
larger than the Bloch period allows resolved intersite transitions
$\left|g,m\right\rangle \rightarrow \left|e,m\pm\Delta
m\right\rangle$, at Raman frequencies :$$\nu_{R}=\nu_{HFS}\pm
\Delta m\times\nu_{B}$$ where $\nu_{HFS}$ is the hyperfine
splitting and $\Delta m$ is the number of lattice wells separating
the two coupled WS states. The energy spectrum of our system is
schematically illustrated in Fig. \ref{fig:energy}.

\begin{figure}
    \centering
        \includegraphics[scale=0.42]{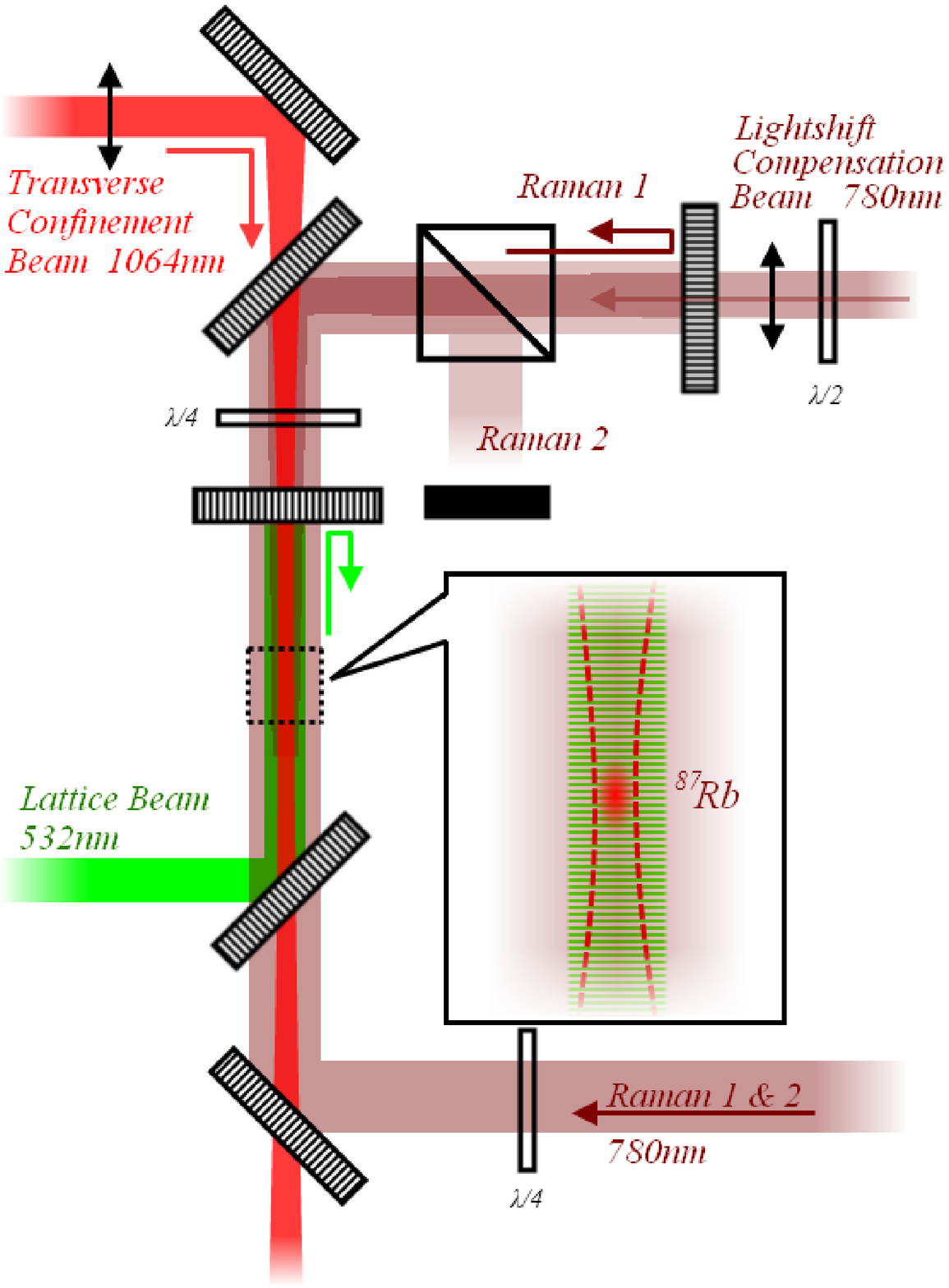}
   
    \caption{Experimental setup for the optical trapping and Raman intersite transitions.
The different beams are superposed using dichroic mirrors. The
Raman beams are also superposed and one of them is retro-reflected
to allow counterpropagating transitions.}
 \label{setup}
\end{figure}

Coupling between neighboring wells can be efficiently tuned using
the lattice depth when $k_{l}$ is close to $k_{eff}$, where
$k_{l}$ is the optical lattice wave vector \cite{pereiraFG}. We
therefore use a mixed trap configuration with a blue detuned
lattice generated by a single mode frequency doubled Nd:YVO$_{4}$
laser ($\lambda_{l}=532$ nm, beam waist $600$ $\mu$m) that
provides only vertical longitudinal confinement, superposed with a
red detuned ($\lambda = 1064$ nm, beam waist $200$ $\mu$m ) Yb
fiber laser providing transverse confinement (see Fig.
\ref{setup}). To load this dipole trap, we first accumulate up to
$10^{7}$ atoms in a 3D-Magneto-Optical trap (MOT) fed by a 2D-MOT. The cloud is then
cooled down to $2$ $\mu$K by a far detuned molasses,
at the end of which we switch off
the lasers to let the untrapped atoms fall. At our low lattice
depth ($U_{l}\simeq 4E_{R}$ (where
$E_{R}=(\hbar k_l)^2/(2 m_a)$ is the lattice recoil energy), only
the first band has a non-negligible lifetime and is populated with
about $10^{5}$ atoms vertically distributed along $10^{4}$ sites
(the second band is centered at $5 E_{R}$ already above the
lattice depth). The atoms accumulated in all the Zeeman sublevels
of $\left|5^{2}S_{1/2},F=2\right\rangle$ are depumped to
$\left|5^{2}S_{1/2},F=1\right\rangle$ and then optically pumped
($95 \%$ efficiency) on the $\left|5^{2}S_{1/2},F=1\right\rangle
\rightarrow \left|5^{2}P_{3/2},F=0\right\rangle$ transition to the
$\left|5^{2}S_{1/2},F=1,m_{F}=0\right\rangle$ Zeeman sublevel,
which is sensitive to stray magnetic fields only to second order.
The remaining $5 \%$ unpolarized atoms can easily be removed from
the trap with a pushing beam. Our fluorescence detection scheme,
based on a time of flight measurement similar to the one used in
atomic clocks and inertial sensors, allows us to measure the
atomic populations in the two hyperfine states after releasing the
atoms from the trap \cite{detection}. The Raman transitions are
driven by two counterpropagating beams at $780$ nm circularly
polarized, detuned from the atomic transition by about $3$ GHz,
and aligned along the direction of the optical trap beams. The
beams are collimated with a $1/e^{2}$ radius of $1$ cm, ensuring a
good intensity homogeneity along the transverse size of the trap
(about $200$ $\mu$m radius).

\begin{figure}
    \centering
        \includegraphics[scale=0.6]{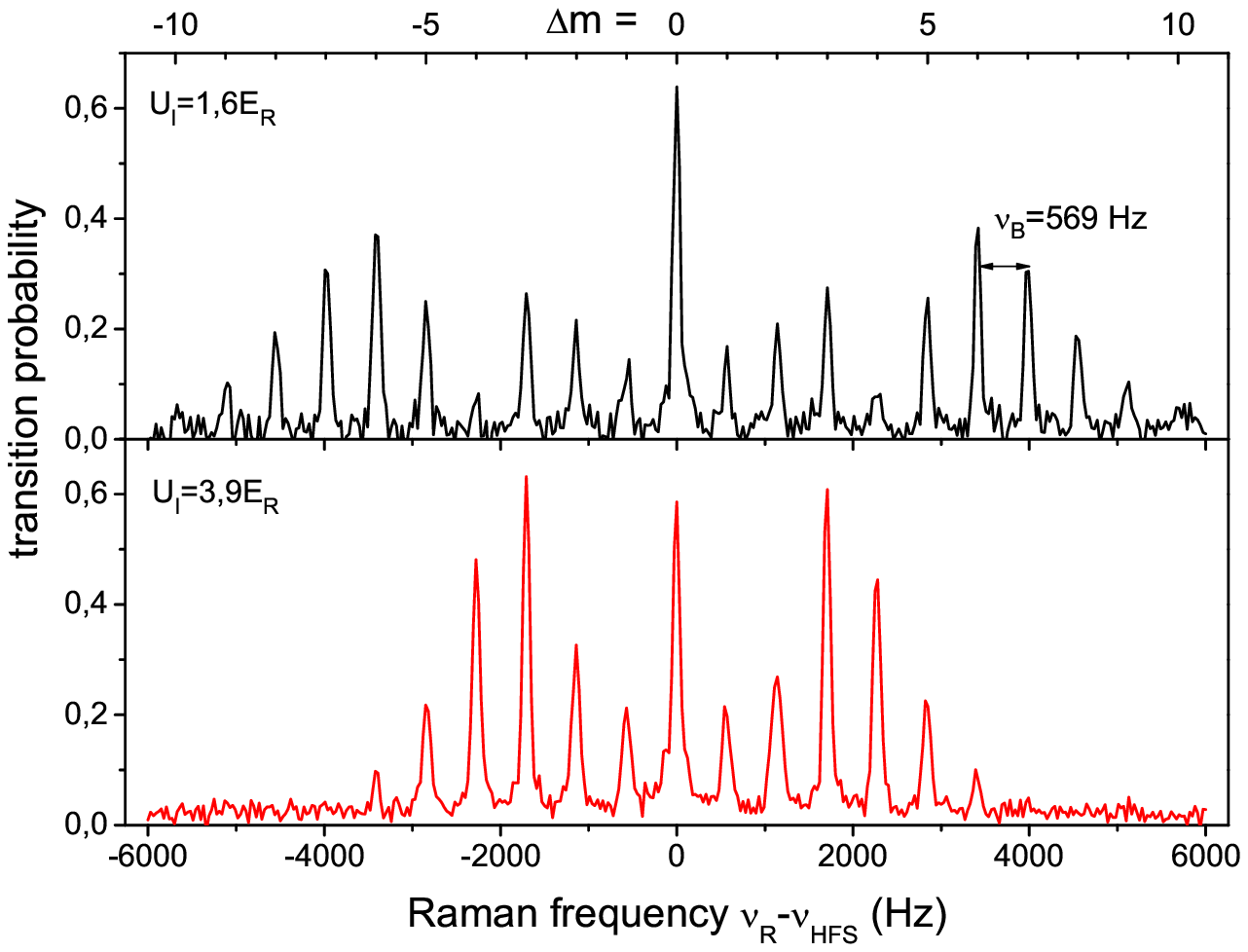}
  
    \caption{Raman spectra for two different lattice depths, showing evidence of transitions
between up to 9 neighboring lattice sites, each having a different
Rabi frequency according to equation \ref{eq:1}. The excitation
time is 10 ms, which is smaller than the duration of a $\pi$ pulse
for each transition. The peaks are separated by the Bloch
frequency of our system $\nu_{B}\approx 569$ Hz.}
  \label{spectrum}
\end{figure}

Fig. \ref{spectrum} shows two typical Raman spectra of the
transition probability as a function of the Raman frequency
$\nu_{R}$, taken for two different lattice depths. Transitions
between the two hyperfine levels at Raman frequencies equal to the
hyperfine splitting plus or minus an integer number $\Delta m$ of
Bloch frequencies ($\nu_{B}\approx 569$ Hz in our system) are the
signature that the atoms actually tunneled across $\Delta m$
lattice sites. For those scans, the intensities in the Raman laser
beams were $0.25$ and $0.54$ mW$/$cm$^{2}$. The resulting Rabi
frequencies $\Omega_{\Delta m}$, different for each transition,
are always smaller than the Bloch frequency, so that each peak is
well resolved. The ratio between the Raman intensities was chosen
to cancel the differential light shift of the hyperfine transition
induced by them \cite{Raman}. The Rabi frequency for each
transition $\Delta m$ is written \cite{coupling}:

\begin{equation}
\Omega_{\Delta m}=\Omega_{U_{l}=0}\left\langle W_{m}\right|e^{-ik_{eff}x}\left|W_{m\pm\Delta m}\right\rangle
\label{eq:1}
\end{equation}

where $\Omega_{U_{l}=0}$ is the Rabi frequency in free space. Due
to the translational symmetry of the WS states, $\Omega_{\Delta
m}$ doesn't depend on the initial well index $m$ but only on the
absolute value of $\Delta m$ \cite{coupling}. It also depends on
the lattice wavelength $\lambda_{l}$ and depth $U_{l}$, which is
an important feature of this experiment, as it induces a spatial
inhomogeneity on the Rabi frequency seen by the trapped atoms via
the transverse inhomogeneity of the lattice depth in the trap. The
damping induced on the Rabi oscillations by this inhomogeneity is
the main limitation on the transfer efficiency of the Raman
transitions.

  \begin{figure}
    \centering
        \includegraphics[scale=0.55]{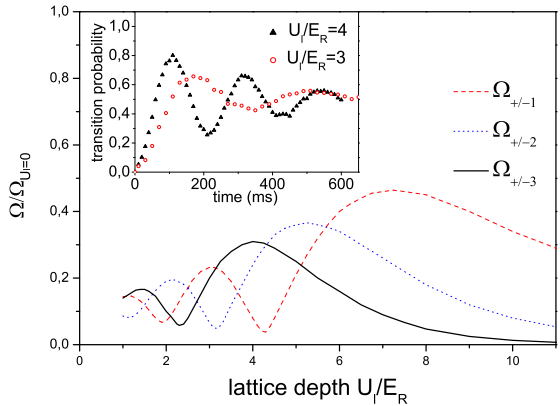}
   
    \caption{Calculation of the normalized Rabi frequencies for $\Delta m = \pm 1$, $\Delta m = \pm 2$ and $\Delta m = \pm 3$ transitions, as a function of the lattice depth. Inset : Experimental Rabi oscillations on the transition $\Delta m=-3$, for $U_{l}=4E_{R}$ and $U_{l}=3E_{R}$.}
     \label{rabi}
\end{figure}

We calculated $\Omega_{\Delta m}$ for the parameters
of our system, as a function of $U_{l}$ and for various values of
$\Delta m$. The result is shown in Fig. \ref{rabi} for $\Delta m
= \pm 1$, $\Delta m = \pm 2$ and $\Delta m = \pm 3$. To limit the
lattice depth inhomogeneity due to the transverse extension of the
atomic cloud, the $1/e^{2}$ waist of the beam providing transverse
confinement is smaller than the one of the lattice beam. Moreover,
we can choose to tune the lattice depth at a value where, for a
transition of interest, the variation of the coupling with lattice
depth $ \Delta \Omega(U_{l}) / \Delta U_{l}$ is small, as
illustrated by the inset of Fig. \ref{rabi}. On this graph, we
compare the shape of Rabi oscillations at resonance for the
$\Delta m = -3$ transition and for two different lattice depths.
Besides a difference in the period of the Rabi oscillations,
we observe that the best contrast is obtained for $U_{l}\simeq 4
E_{R}$, where the coupling inhomogeneity is lower, allowing us to
reach a transfer efficiency of about 80 $\%$. The lattice depth is
estimated by measuring the Rabi frequencies for different
transitions and comparing them to the calculation.

\begin{figure}
    \centering
        \includegraphics[width=8 cm]{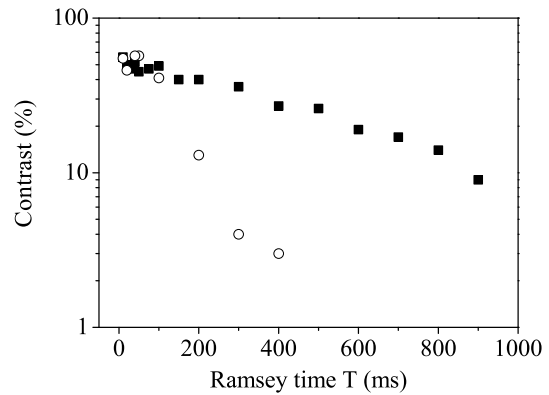}
  
    \caption{Contrast of the interferometer versus Ramsey interrogation time $T$. Open circles (resp. black squares) display the contrast without (resp. with) the light shift compensation beam.}
    
      \label{contrast}
\end{figure}

We investigated the question of the coherence time of the trapped states. As a diagnosis tool, we performed Ramsey spectroscopy on the $\left|g,m\right\rangle \rightarrow \left|e,m+3\right\rangle$ transition. As there is no initial atomic coherence from one site to it's neighbors, the atoms distributed in many lattice sites can be treated as independent interferometers. The phase is read out by the measurement of the internal atomic state population. Our interferometer consists in two Raman $\pi /2$ pulses of frequency $\nu_{R}$ scanned close to $\nu_{HFS}+3\nu_{B}$, separated by a time interval $T$. The intensity of the Raman lasers was chosen for the Rabi frequency $\Omega_{\Delta m=3}/2\pi$ to be much smaller than $\nu_{B}$, in order to ensure a good separation with neighboring transitions. The duration of the $\pi/2$ pulses is then $\tau_{\pi/2}=5.5$ ms. 
Fig. \ref{contrast} shows as open circles the evolution of the
contrast as we increase the interrogation time $T$. The contrast
at short $T$ is about $65 \%$, which is coherent with the $80 \%$
efficiency of each of the two Raman $\pi /2$ pulses. When
increasing $T$, the contrast decreases rapidly, which is due to
the transverse inhomogeneity of the differential light shift
induced on the hyperfine transition by the Gaussian profile of the
IR laser used for transverse trapping \cite{meschede2}. This
position dependent frequency shift induces an inhomogeneous
broadening along the transverse direction, which is the main
effect limiting the coherence lifetime in the lattice trap.

Many schemes have been proposed and demonstrated to cancel this
source of inhomogeneous dephasing in an optical trap
\cite{cancel2,cancel1,cancel3}. One particularly efficient method
in our case is to add a low power laser beam, mode matched with
the transverse trapping beam, and whose frequency is tuned between
the two hypefine levels, as reported in \cite{kaplan}. This beam
compensates the differential ligthshift induced by the transverse
trapping light with a laser power of only a few tens of nW, so
that decoherence due to photon scattering is negligible at the
experiment's time scale. In practice, we use as a compensating
beam a fraction of one of the two Raman beams, with an additional
detuning of 80 MHz in order to prevent undesired Raman
transitions. This beam is superposed with the IR laser (see Fig. \ref{setup}), and its
size, position and power are adjusted to optimize the contrast of
the interferometer for long interrogation times of several
hundreds of milliseconds, for which otherwise the contrast is
zero. In our case, for an IR power of about 2 W, the differential
light shift is compensated with a power of $12$ nW. Fig.
\ref{contrast} displays as black squares the evolution of the
contrast vs T with the compensating beam, and clearly shows the
improvement of the lifetime of the coherence.

Finally, as a preliminary investigation, we evaluated the
frequency sensitivity of the interferometer. This is
motivated by the possibility to use this trapped atomic spatial
interferometer for metrology applications such as gravimetry or
short range forces measurement \cite{wolf}. We locked the Raman
lasers frequency difference to the center fringe of the
interferometer, with a computer controlled servo integrator.
Performing the measurement alternatively on the left and right
transitions ($\Delta m=\pm3$) and calculating the difference of
the measured frequencies allows us to measure $6\nu_B$ while
cancelling the frequency shifts of the hyperfine clock frequency,
due to, for example, the quadratic zeeman effect and the
differential lightshifts induced by the trapping lasers. For an
interrogation time of $T=400$ ms and a cycle time of $T_c=1.4$ s,
the Allan standard deviation of the frequency difference decreases
as $0.1$ Hz$.\tau^{-1/2}$ with $\tau$ the integration time in
seconds. This corresponds to a statistical uncertainty on the
measurement of the Bloch frequency of $6\times10^{-5}$ in relative value
after $1$ s integration.

Our Wannier-Stark interferometer shows great potential for metrology applications. As an example, we plan to perform it close
to the reflecting surface of the lattice, which would allow the
measurement of short range forces (Casimir-Polder, short range
modifications of gravity)\cite{pereiraFG}. The statistical uncertainty in the
measurement of the Casimir-Polder potential, assuming the
performance demonstrated here, would reach 1$\%$ for a distance of
$5$ $\mu$m and a measurement time of $1000$ s.

The technique for controlled and coherent transport of atoms
demonstrated in this work is unique in terms of versatility. The
high resolution reachable by the Raman transitions (up to 1 Hz)
suggests the possibility of selectively addressing one single
lattice site, using for example the lightshift induced by a
focused laser to lift the degeneracy between the transitions.
Although demonstrated here for a thermal cloud, this technique is
perfectly suitable for degenerate quantum gases. Besides, it also works with one photon transitions, as  recently highlighted in \cite{bize}. All these
features make this tool a potential candidate for the realization
of quantum logic operations.

\begin{acknowledgments}
This research is carried on within the project iSense, which acknowledges the financial support of the Future and Emerging Technologies (FET) programme within the Seventh Framework Programme for Research of the European Commission, under FET-Open grant number: 250072. We also gratefully acknowledge support by Ville de Paris (Emergence(s) program) and IFRAF. G.T. thanks the Intercan network and the UFA-DFH for financial
support.

Helpful discussions with A. Landragin, P. Lemonde, A. Clairon, S. Bize, M-C. Angonin and R. Messina are greatfully acknowleged.

\end{acknowledgments}


\begin{thebibliography}{30}

\bibitem{cirac} J. J. Garcia-Ripoll, M. A. Martin-Delgado, and J. I. Cirac,
Phys. Rev. Lett. \textbf{93}, 250405 (2004).

\bibitem{latticeclock} M. Takamoto, F.-L. Hong, R. Higashi, and H. Katori,
Nature (London) \textbf{435}, 321 (2005)

\bibitem{hsurm} P. Clad\'{e}, E. de Mirandes, M. Cadoret, S. Guellati-Kh\'{e}lifa, C. Schwob, F. Nez, L. Julien, and F. Biraben, Phys. Rev. Lett. \textbf{96}, 033001 (2006)


\bibitem{qip} H.J. Briegel, T. Calarco, D. Jaksch, J. I. Cirac and P. Zoller, J. Mod. Opt. \textbf{47}, 415–451 (2000).


\bibitem{single} D. Schrader, I. Dotsenko, M. Khudaverdyan, Y. Miroshnychenko, A. Rauschenbeutel, and D. Meschede, Phys. Rev. Lett. \textbf{93}, 150501 (2004)

\bibitem{ql} G. K. Brennen, C. M. Caves, P. S. Jessen, and I. H. Deutsch, Phys. Rev. Lett. \textbf{82}, 1060–1063 (1999)

\bibitem{microwave} L. F\"orster, M. Karski, J. Choi, A. Steffen, W. Alt, D. Meschede, and A. Widera, E. Montano, J. Hoon Lee, W. Rakreungdet, and P. S. Jessen, Phys. Rev. Lett. \textbf{103}, 233001 (2009)

\bibitem{salomon} M. Ben Dahan, E. Peik, J. Reichel, Y. Castin, and C. Salomon,
Phys. Rev. Lett. \textbf{76}, 4508 (1996)

\bibitem{raizen} S. R. Wilkinson, C. F. Bharucha, K.W. Madison, Q. Niu,
and M. G. Raizen, Phys. Rev. Lett. \textbf{76}, 4512 (1996)

\bibitem{tino} V.V. Ivanov, A. Alberti, M. Schioppo, G. Ferrari, M. Artoni, M. L. Chiofalo and G. M. Tino, Phys. Rev. Lett. {\bf 100}, 043602 (2008)

\bibitem{Bloch} O. Mandel, M. Greiner, A. Widera, T. Rom,
T. W. H\"ansch and I. Bloch, Nature (London) \textbf{425}, 937 (2003)

\bibitem{Porto} M. Anderlini, P. J. Lee, B. L. Brown, J. Sebby-Strabley, W. D. Phillips and J. V. Porto, Nature (London) \textbf{448}, 452 (2007)

\bibitem{pereiraFG} F. Pereira dos Santos et al., \textit{Proceedings of the $7^{th}$ symposium Frequency Standards and Metrology}, (World scientific printers, Singapore, 2009)

\bibitem{detection} J. Le Gou\"et, T. E. Mehlst\"aubler, J. Kim, S. Merlet, A. Clairon,
A. Landragin, and F. Pereira Dos Santos, Appl. Phys. B \textbf{92}, 133
(2008)

\bibitem{Raman} D.S. Weiss, B.C. Young and S. Chu, Applied Physics B \textbf{59}, 217-256 (1994)

\bibitem{coupling} P. Lemonde and P. Wolf, Phys. Rev. A \textbf{72}, 033409 (2005)



\bibitem{meschede2} S. Kuhr, W. Alt, D. Schrader, I. Dotsenko, Y. Miroshnychenko, A. Rauschenbeutel and D. Meschede, Phys. Rev. A {\bf 72}, 023406 (2005)

\bibitem{cancel2} M. F. Andersen, A. Kaplan, and N. Davidson,  Phys. Rev. Lett. \textbf{90}, 023001 (2003)

\bibitem{cancel1} A. Derevianko Phys. Rev. Lett. \textbf{105}, 033002 (2010)

\bibitem{cancel3} R. Chicireanu, K. D. Nelson, S. Olmschenk, N. Lundblad, A. Derevianko, and J. V. Porto, Phys. Rev. Lett. \textbf{106}, 063002 (2011)



\bibitem{kaplan} A. Kaplan, M. F. Andersen, and N. Davidson, Phys. Rev. A \textbf{66}, 045401 (2002)

\bibitem{wolf} P. Wolf, P. Lemonde, A. Lambrecht, S. Bize, A. Landragin, and A. Clairon, Phys. Rev. A \textbf{75}, 063608 (2007)

\bibitem{bize} L. Yi, S. Mejri, J. J. McFerran, Y. Le Coq, and S. Bize, Phys. Rev. Lett. \textbf{106}, 073005 (2011)


\end{thebibliography}

\end{document}